\documentclass{article}
\usepackage{ijcai19}

\usepackage{times}
\usepackage{soul}
\usepackage{url}
\usepackage[hidelinks]{hyperref}
\usepackage[utf8]{inputenc}
\usepackage[small]{caption}
\usepackage{graphicx}
\usepackage{amsmath,amssymb,amsfonts}
\usepackage{booktabs}
\usepackage{algorithm}
\usepackage{algorithmic}
\title{Fast Multi-resolution Segmentation for Nonstationary Hawkes Process Using Cumulants}

\author{
Feng Zhou$^{1,2}$\and
Zhidong Li$^3$\and
Xuhui Fan$^2$\and
Yang Wang$^3$\and
Arcot Sowmya$^2$\and
Fang Chen$^3$\\
\affiliations
$^1$Data61 CSIRO\\
$^2$University of New South Wales\\
$^3$University of Technology Sydney\\
}

\begin{document}

\maketitle

\begin{abstract}
The stationarity is assumed in vanilla Hawkes process, which reduces the model complexity but introduces a strong assumption. In this paper, we propose a fast multi-resolution segmentation algorithm to capture the time-varying characteristics of nonstationary Hawkes process. The proposed algorithm is based on the first and second order cumulants. Except for the computation efficiency, the algorithm can provide a hierarchical view of the segmentation at different resolutions. We extensively investigate the impact of hyperparameters on the performance of this algorithm. To ease the choice of one hyperparameter, a refined Gaussian process based segmentation algorithm is also proposed which proves to be robust. The proposed algorithm is applied to a real vehicle collision dataset and the outcome shows some interesting hierarchical dynamic time-varying characteristics. 
\end{abstract}

\section{Introduction}
\label{sec1}
The point process data is a common data type in real applications. To model this kind of point process data, various statistical models have been proposed to disclose its underlying temporal dynamics, such as homogeneous Poisson process \cite{thompson2012point}, inhomogeneous Poisson process \cite{weinberg2007bayesian} and Hawkes process \cite{hawkes1971spectra}. In this paper, we focus on Hawkes process. 

Hawkes process is widely used to model the self-exciting phenomenon which can be observed in many fields, like crime \cite{liu2018exploiting}, ecosystem \cite{gupta2018discrete}, transportation \cite{du2016recurrent} and TV programs \cite{luo2015multi}. An important way to characterize a temporal point process is through the definition of a conditional intensity. The specific Hawkes process conditional intensity is:
\begin{equation}
\lambda(t)=\mu+\int_0^t\phi(t-s)d\mathbb{N}(s)=\mu+\sum_{t_i<t}\phi(t-t_i),
\label{1}
\end{equation}
where $\mu>0$ is the baseline intensity which is constant, $\{t_i\}$ are the timestamps of events before time $t$, $\mathbb{N}(t)$ is the corresponding counting process and $\phi(\cdot)$ is the triggering kernel. The summation of triggering kernels explains the nature of self-excitation, which is the occurrence of events in the past will intensify the intensity of events occurring in the future. 

\begin{figure}[t]
\centerline{\includegraphics[width=0.5\textwidth]{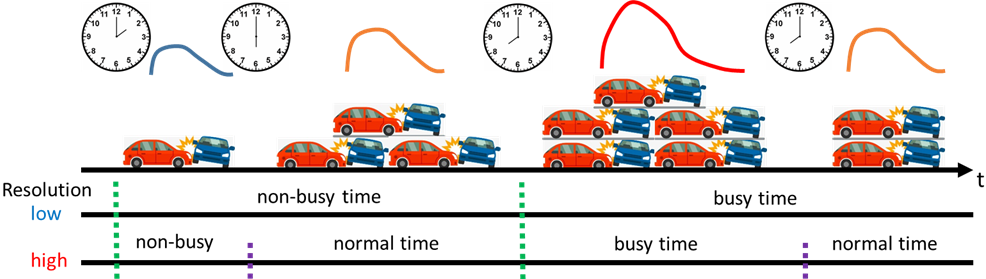}}
\caption{The multi-resolution segmentation of triggering effect of vehicle collisions. The low-resolution (2 segments) and high-resolution (4 segments) partitions provide a hierarchical insight into the dynamic time-varying characteristics.}
\label{fig2}
\end{figure}

It is straightforward to see that the conditional intensity of Hawkes process is unchanged over timeshifting because $\mu$ is a constant and $\phi(\cdot)$ only depends on $\tau=t-t_i$ not on $t$, which means stationarity. The assumption of stationarity leads to reduced model complexity and easy inference. However, the point process data generated in many real applications has nonstationary properties, which means its first, second and higher order cumulants (moments) are changing over time. Applying the vanilla Hawkes process directly to the nonstationary data is apparently inappropriate. On the other hand, the nonstationarity itself can be an important feature in some applications. For example, 
in transportation, the influence of a car accident to the road condition is changing between day and night and between busy and non-busy hours (see Fig.~\ref{fig2}).

One of the common methods of analyzing nonstationary time series is to use segmentation. This kind of problem is also called change-point problem which is studied in mathematics \cite{carlstein1994change}. Given a nonstationary point process data, the segmentation algorithm will divide the whole observation period into several non-overlapping contiguous segments in such a way that each segment is more approximately stationary than the original data and can be assumed to be stationary. 

To the best of our knowledge, no segmentation algorithm has been proposed for nonstationary Hawkes process. In this paper, we propose the first multi-resolution segmentation (MRS) algorithm for the nonstationary Hawkes process which can reveal the optimal partition structure in a hierarchical manner. The multi-resolution segmentation is meaningful in real data applications. For example, when the traffic data is analyzed (see Fig.~\ref{fig2}), the low-resolution partition (e.g. two segments) corresponds to a ``coarser" distinction (e.g. day and night), while the high-resolution partition (e.g. four or six segments) corresponds to a ``finer" distinction (e.g. the alternating busy and non-busy hours). This will help us obtain a hierarchical insight into the nonstationary structure of point process data. 


As shown later, the performance of the MRS algorithm depends on the choice of hyperparameters. To ease the choice of one hyperparameter, we propose a revised Gaussian process based version which is more robust. Overall, our work makes the following contributions: 
\begin{itemize}
\item We propose the first multi-resolution segmentation algorithm which provides a hierarchical analysis of the dynamic evolution of nonstationary Hawkes process. 
\item The MRS depends on the cumulants of Hawkes process which is fast to compute. Consequently, the MRS (linear computation complexity) is faster than the case of direct estimation of baseline intensity and triggering kernel. 
\item A more robust revised version of MRS is also proposed to ease the choice of one hyperparameter, which is slower but still acceptable in real applications. 
\end{itemize}


\section{Related Works}
\label{sec2}



\subsection{Nonstationary Hawkes Process}

The relaxation of vanilla Hawkes process to nonstationary version mainly consists of two cases: the first case is the extension of baseline intensity $\mu$ to time-changing $\mu(t)$ and the second case is the extension of triggering kernel $\phi(\tau)$ to time-changing $\phi(\tau,t)$. Plenty of state of the arts have performed inference for a time-changing baseline intensity with a stationary triggering kernel \cite{lewis2011nonparametric,lemonnier2014nonparametric,tannenbaum2017theory}. For both baseline intensity and triggering kernel being nonstationary, \cite{roueff2016locally} and \cite{roueff2017time} provided a general nonparametric estimation theory for the first and second order cumulants of a locally stationary Hawkes process. This method is general but low in computation efficiency: because every point on the two dimensional covariance function $Cov(\tau,t)$ has to be estimated, it is not applicable to real applications. In this sense, the MRS algorithm can be considered as a ``coarser" version of \cite{roueff2016locally}: it combines adjacent small sectors with similar statistical properties into a larger segment and only outputs more heterogeneous segments. Although it is ``coarser", the computation complexity is drastically reduced to make it practical. 

\subsection{Segmentation of Time Series}

Segmentation is a standard method of data analysis to divide a nonstationary time series into a certain number of non-overlapping contiguous homogeneous segments. A heuristic segmentation algorithm is desinged to study the distribution of periods with constant heart rate in \cite{bernaola2001scale}. The same method is also applied to analyze changes of the climate \cite{feng2005abrupt}. A generalized version is proposed in \cite{bernaola2012segmentation} to overcome the oversegmentation problem caused by heterogeneities induced by correlations. Similarly, \cite{toth2010segmentation} generalizes this existing algorithm for segmenting regime switching processes. All segmentation methods mentioned above cannot be applied directly to Hawkes process, because they only consider the case of (marked) Poisson process. 

\section{Cumulants of Hawkes Process}
\label{sec3}

The cumulants of Hawkes process \cite{bacry2016first,jovanovic2015cumulants} are used in the MRS algorithm. We consider a 1-variate Hawkes process $\mathbb{N}_t$ whose jumps are all of size 1 and whose intensity at time $t$ is $\lambda(t)$. If $\{t_i\}$ denotes the jump times of $\mathbb{N}_t$, the $\lambda(t)$ can be expressed as \eqref{1}. If $\mathbb{N}_t$ is stationary, the first order cumulant (mean event rate) is
\begin{equation}
\Lambda dt=\mathbb{E}(d\mathbb{N}_t)=\frac{\mu}{1-\int\phi(\tau)d\tau}dt.
\label{2}
\end{equation}
The second order cumulant is 
\begin{equation}
Cov(d\mathbb{N}_{t_1},d\mathbb{N}_{t_2})=\mathbb{E}(d\mathbb{N}_{t_1}d\mathbb{N}_{t_2})-\mathbb{E}(d\mathbb{N}_{t_1})\mathbb{E}(d\mathbb{N}_{t_2}).
\label{3}
\end{equation}
Because $\mathbb{N}_t$ is stationary, $Cov(d\mathbb{N}_{t_1},d\mathbb{N}_{t_2})$ only depends on $\tau=t_2-t_1$ and can be expressed as: 
\begin{equation}
v(\tau)d\tau=\mathbb{E}(d\mathbb{N}_{0}d\mathbb{N}_{\tau})-\mathbb{E}(d\mathbb{N}_{0})\mathbb{E}(d\mathbb{N}_{\tau}).
\label{4}
\end{equation}
Or, it can be rewritten in terms of conditional expectations
\begin{equation}
g(\tau)d\tau=v(\tau)d\tau/\Lambda=\mathbb{E}(d\mathbb{N}_{\tau}|d\mathbb{N}_0=1)-\Lambda d\tau.
\label{5}
\end{equation}
The $g(\tau)$ will be used throughout this paper. 

A stationary Hawkes process is uniquely defined by its first and second order cumulants and there is a bijection between its second order statistics $g(\tau)$ and the triggering kernel $\phi(\tau)$. 

\section{Multi-resolution Segmentation}
\label{sec4}

We assume there is a set of observation $\{t_i\}_{i=1}^N$ on $[0,T]$ from a nonstationary Hawkes process where the baseline intensity $\mu$ is piecewise constant and the triggering kernel $\phi(\tau)$ is changing over time $t$. Given $M$, the fundamental idea of MRS is to uniformly divide the observation period $[0,T]$ into $M$ sectors (the highest resolution), e.g. $s_1,s_2,...,s_M$, where $\{s_j\}_{j=1}^M$ are sectors and $|s_j|$ is the width of the sector. In each $s_j$, the point process is assumed to be stationary. 

Intuitively, we can estimate the triggering kernel $\phi(\tau)$ in each sector, compare them by adjacent pairs, and find out the most possible partition positions. However, the estimation of $\phi(\tau)$ is time consuming no matter in parametric way (maximum likelihood) or nonparametric way (EM algorithm \cite{lewis2011nonparametric}, Wiener-Hopf equation \cite{bacry2016first}), let alone running on all sectors. In order to increase computation efficiency, we do not estimate $\phi(\tau)$ in each sector directly but use the second order statistics $g_j(\tau)$ instead which can be estimated faster. The second order statistics $g_j(\tau)$ in each sector can be empirically estimated using the empirical version of \eqref{5}. 

The reason we can replace $\phi(\tau)$ in each sector with $g_j(\tau)$ is that there is a bijection between them, so the difference between two adjacent $g_j(\tau)$ stands for the nonstationarity of $\phi(\tau)$. The difference of two adjacent $g_j(\tau)$ is written as a normalized mean squared error (NMSE)
\begin{equation}
NMSE=\mathbb{E}_{\tau}\left((\frac{g_j(\tau)}{\int{g_j(\tau)}d\tau}-\frac{g_{j+1}(\tau)}{\int{g_{j+1}(\tau)}d\tau})^2\right).
\label{6}
\end{equation}

In most cases, $g_j(\tau)$ is an even function for 1-variate Hawkes process when $\tau\to\pm\infty$, $g_j(\tau)\to 0$. If $g_j(\tau)$ is expressed as a histogram function $g_j(\tau)=\sum_{k=1}^K(g_j^k\delta_{kh})$ where $\delta_{kh}(\tau)=1$ if $(k-1)h\leq\tau<kh$ and $0$ otherwise, $h$ is the bin-width and $g_j(\tau)$ is $0$ beyond the support of $Kh$, we can write $g_j(\tau)$ as a vector $\mathbf{g}_j=[g_j^k]_{k=1}^K$. Equation \eqref{6} can be converted to a discrete version
\begin{equation}
NMSE=\frac{\sum_{k=1}^K\left((\frac{g_j^k}{2h\sum_{k=1}^Kg_j^k}-\frac{g_{j+1}^k}{2h\sum_{k=1}^Kg_{j+1}^k})^2\right)}{K}. 
\label{7}
\end{equation}

\begin{figure}[htbp]
\centerline{\includegraphics[width=0.44\textwidth]{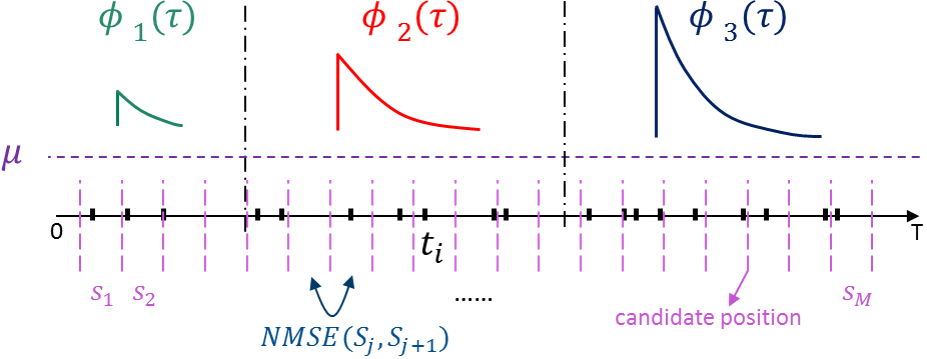}}
\caption{The scheme of multi-resolution segmentation, for simplicity $\mu$ is assumed to be constant and there are 3 different $\phi(\tau)$'s distributed on $[0,T]$.}
\label{fig1}
\end{figure}

Given the NMSE on all candidate cutting positions, if a desired number of segments (the desired output resolution) $R$ is set, we can pick out the largest $R-1$ cutting positions which is the segmentation. 
The scheme of MRS is shown in Fig.~\ref{fig1}. By multi-resolution, we mean by increasing (decreasing) the desired output resolution $R$ the segmentation algorithm will output segments at different resolutions in a hierarchical manner. For example, when $R=M$, the partitioner will output the highest resolution (cutting at all candidate positions), as $R$ becomes smaller the output resolution will be lower (fewer segments will be given out) until there is no cutting at all.

After segmentation, we can piecewisely learn the baseline intensity and triggering kernel on each segment. Specifically, a nonparametric estimation method: Wiener-Hopf equation method \cite{bacry2016first} is used. As proved in that work, $\phi(\tau)$ and $g(\tau)$ satisfy the Wiener-Hopf equation
\begin{equation}
g(\tau)=\phi(\tau)+\phi(\tau)*g(\tau), \forall{\tau>0},
\label{8}
\end{equation}
where $*$ stands for convolution. In most cases, the Winer-Hopf equation cannot be solved analytically, but there is a lot of literature \cite{noble1959methods,atkinson1976survey} on how to solve it numerically. A common method is the Nystrom method \cite{nystrom1930praktische}. After the solution of $\phi(\tau)$, $\mu$ can be estimated by the first order cumulant \eqref{2}. 

\section{Synthetic Data Experiment}
\label{sec5}

We use thinning algorithm \cite{ogata1998space} to independently generate 40 sets of observations $\{\{t_i\}_{i=1}^{N_l}\}_{l=1}^{40}$ ($N_l$ is the number of points on $l$-th observation) on $[0,1000]$ from a nonstationary Hawkes process where $\mu$'s are $2$, $1.5$, $1$ and triggering kernels are $\phi_1(\tau)=1\cdot\exp(-2\tau)$, $\phi_2(\tau)=2\cdot\exp(-4\tau)$ and $\phi_3(\tau)=3\cdot\exp(-4\tau)$ distributed on $[0,200]$, $[200,600]$ and $[600,1000]$, respectively (see Fig.~\ref{fig1}). The goal is to find the underlying partition structure and estimate $\mu$'s and $\phi(\tau)$'s in a nonparametric way. The highest resolution is set to be $M=10$ ($|s_j|=100$), $g_j(\tau)$ is expressed as a histogram function $g_j(\tau)=\sum_{k=1}^K(g_j^k\delta_{kh})$ where $h=0.75$ and $K=8$. 


We average the estimated $g_j(\tau)$ over 40 sets of independent observations and Tab.~\ref{tab2} shows the multi-resolution segmentation results in a hierarchical manner as $R$ increases from $1$ to the highest resolution $10$. We can see when $R=1$, there is no cutting at all; when $R=3$, the partition positions match with the ground truth; when $R=10$ the algorithm cuts at every candidate position (the highest resolution). To quantify the NMSE caused by estimation variance, the proportion of the minimum threshold corresponding to $R$ over the maximum NMSE is shown in Tab.~\ref{tab2}. We can see the NMSE induced by estimation variance is below 11.41\% (the last correct cutting is ``200" which corresponds to 11.41\%), which means the MRS is robust to obtain the correct segmentation. 

\begin{table}[htbp]
\centering
\scalebox{0.74}{
\begin{tabular}{|c|c|c|c|c|c|}
\hline
\textbf{$R$} & 1             & 2       & 3   & 4 & 5\\ \hline
New Position & $\varnothing$ & 600 & 200 & 500 & 900\\ \hline
\textbf{$\frac{\text{Min(Threshold)}}{\text{Max(NMSE)}}$} & 100\% & 88.45\% & 11.41\% & 8.05\% & 7.15\%\\ \hline
\textbf{$R$} & 6 & 7 & 8 & 9 & 10\\ \hline
New Position & 400 & 300 & 700 & 800 & 100\\ \hline
\textbf{$\frac{\text{Min(Threshold)}}{\text{Max(NMSE)}}$} & 6.88\% & 5.17\% & 2.24\% & 1.25\% & 0\%\\ \hline
\end{tabular}}
\caption{Multi-resolution segmentation results. ``New Position" is the newly added partition position.}
\label{tab2}
\end{table}

\begin{figure}[htbp]
\centerline{\includegraphics[width=0.5\textwidth]{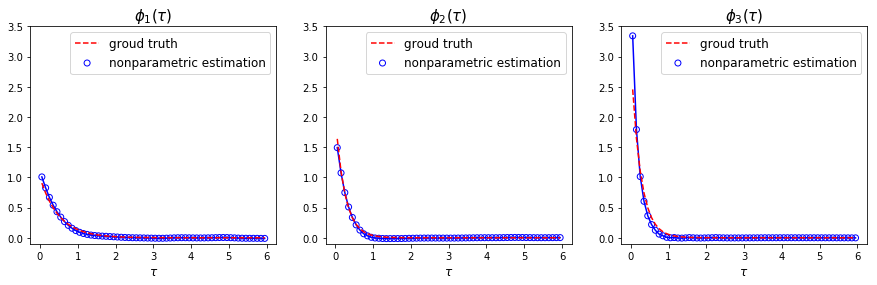}}
\caption{The estimated $\hat{\phi}_1(\tau)$, $\hat{\phi}_2(\tau)$ and $\hat{\phi}_3(\tau)$. The ground truths are $1\cdot\exp(-2\tau)$, $2\cdot\exp(-4\tau)$ and $3\cdot\exp(-4\tau)$, respectively.}
\label{fig4}
\end{figure}

Setting $R=3$, the correct segmentation $[0,200]$, $[200,600]$, $[600,1000]$ is obtained. The next step is to infer $\mu$ and $\phi(\tau)$ on each segment. We empirically estimate the second order statistics $g(\tau)$ on each segment and solve the Winer-Hopf equation \eqref{8}. The estimated $\Hat{\mu}_1=2.05$, $\Hat{\mu}_2=1.64$, $\Hat{\mu}_3=1.01$ and $\Hat{\phi}(\tau)$'s are shown in Fig.~\ref{fig4}. We can see the estimation matches with the ground truth. 

\subsection{Computation Complexity}
\label{c}

In this section, we analyze the computation complexity of MRS algorithm. \textbf{The MRS algorithm has a linear computation complexity which means it is practical}. The complexity of MRS mainly depends on two parameters: the highest resolution $M$ and the size of the observation multiplying the number of bins on $g_j(\tau)$: $\mathcal{N}K$ where $\mathcal{N}=\sum_lN_l$. 

\begin{figure}[htbp]
\centerline{\includegraphics[width=0.37\textwidth]{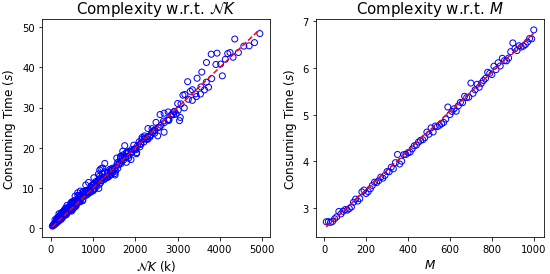}}
\caption{The consuming time of MRS (left: w.r.t. $\mathcal{N}K$ given $M=10$; right: w.r.t. $M$ given $\mathcal{N}K=31,547\times8$).} 
\label{fig5}
\end{figure}

The complexity of estimation of $g_j(\tau)$ on each sector is $\mathcal{O}(n_jK)$ where $n_j$ is the number of points in $s_j$, consequently, the complexity of all $g_j(\tau)$ on all independent observations is $\mathcal{O}(\mathcal{N}K)$. The complexity of NMSE between two adjacent $g_j(\tau)$ over $M$ sectors is $\mathcal{O}(M)$. Therefore, the final complexity of MRS is $\mathcal{O}(\mathcal{N}K+M)$. The time consuming experimental results over $\mathcal{N}K$ ($M$) given $M$ ($\mathcal{N}K$) are shown in Fig.~\ref{fig5} which proves the linear conclusion. The consuming time of estimation of $\phi(\tau)$ and $g(\tau)$ is also compared: given 1,896 observation points, the consuming time of $g(\tau)$ is 0.5 second but 38.4 seconds for $\phi(\tau)$, which proves replacing $\phi(\tau)$ with $g(\tau)$ is more efficient. 

\subsection{Influence of Hyperparameters}

The difference between two adjacent estimated $g_j(\tau)$ and $g_{j+1}(\tau)$ is from two sources: the first source is the difference between $\mathbb{E}(g_j(\tau))$ and $\mathbb{E}(g_{j+1}(\tau))$ which is the nonstationarity, the second source is the estimation variance of $g_j(\tau)$ induced by the choice of hyperparameters. There are two hyperparameters affecting the performance of MRS: $M$ and $K$. 

\subsubsection{Hyperparameter $M$}

Intuitively, the highest resolution $M$ should not be too small or too large. If too small, there are few sectors as the candidate partition positions, consequently, the segmentation result from MRS degrades. If too large, there will be fewer points in each sector $s_j$, which means the estimation variance of $g_j(\tau)$ is large, consequently, the segmentation result also degrades. 

\begin{figure}[htbp]
\centerline{\includegraphics[width=0.5\textwidth]{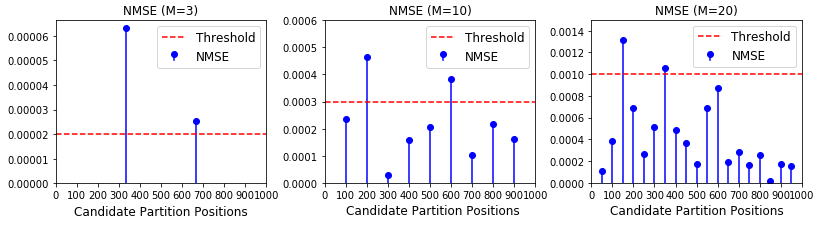}}
\caption{Given $K=8$, the NMSE of MRS w.r.t. $M$. The threshold corresponds to $R=3$ (Only $M=3,10,20$ are shown).}
\label{fig6}
\end{figure}

Given $K=8$, the experiment is performed with $M$ from 3 to 20. The segmentation and NMSE results with $R=3$ are shown in Tab.~\ref{tab3} and Fig.~\ref{fig6}. We can see when $M$ is in $[10,16]$, the segmentation from MRS is close to the ground truth; when $M>20$, the estimation variance is overwhelming, consequently, the partition positions are misidentified. 

\begin{table}
\centering
\scalebox{0.77}{
\begin{tabular}{|c|c|c|c|c|c|}
\hline
\textbf{$M$} & 3        & 10      & 16          & 20    \\ \hline
\textbf{Partition Positions} & 333.3,666.6 & 200,600 & 187.5,687.5 & 150,350 \\ \hline
\textbf{$K$} & 10             & 20       & 30   & 40  \\ \hline
\textbf{Partition Positions} & 200,600 & 200,600 & 100,200 & 100,200 \\ \hline
\end{tabular}}
\caption{Segmentation results of MRS w.r.t. $M$ and $K$}
\label{tab3}
\end{table}

\subsubsection{Hyperparameter $K$} 
\label{db}

Given an appropriate highest resolution $M$, the performance of MRS is also affected by the hyperparameter $K$. The reason behind this phenomenon is that as $K$ becomes larger, there are more bins on $g_j(\tau)$ and the estimated $\mathbf{g}_j=[g_j^k]_{k=1}^K$ will be overfitting. To show this problem, we perform experiments given the highest resolution $M=10$ but with $K=10,40$ and $100$. The estimated $\mathbf{g}_1$ when $K=10,40$ and $100$ is shown in Fig.~\ref{fig8} (only the positive half is shown because of even function). It is clear that the $\mathbf{g}_1$ with $K=100$ is overfitting since there are many spikes up and down. 
 
\begin{figure}[htbp]
\centerline{\includegraphics[width=0.5\textwidth]{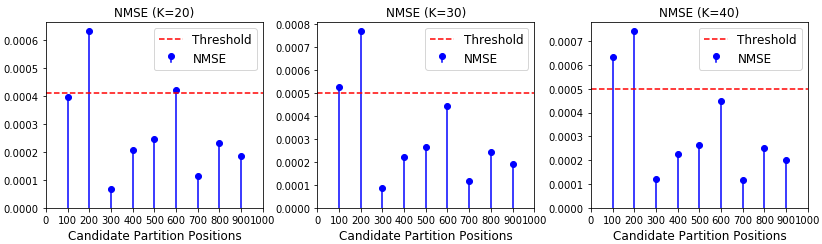}}
\caption{Given $M=10$, the NMSE of MRS w.r.t. $K$. The threshold corresponds to $R=3$ (Only $K=20,30,40$ are shown).}
\label{fig7}
\end{figure}

The more bins we have, the larger the estimation variance of $g_j(\tau)$ will be, which will lead to a misidentified segmentation. To prove it, the segmentation and NMSE results when $K=10,20,30$ and $40$ with $R=3$ are shown in Tab.~\ref{tab3} and Fig.~\ref{fig7}. We can see when $K\geq30$, the segmentation obtained from MRS does not match with the ground truth any more. 


\section{Choice of Hyperparameters}
\label{sec6}

Intuitively, a model selection experiment can be performed to obtain the optimal hyperparameters $M$ and $K$. Nevertheless, for a more robust model, we propose a refined MRS algorithm: GP-MRS in this section, by using which we do not need to choose the optimal value of $K$. We can arbitrarily set a large $K$ as GP-MRS can prevent it from overfitting. 

\subsection{Description of GP-MRS}

The key idea of GP-MRS is to use a standard GP regression to smooth the vector $\mathbf{g}_j=[g_j^k]_{k=1}^K$ in each sector. Given $\mathbf{g}_j=[g_j^1,g_j^2,\dots,g_j^K]$ in $s_j$, the GP regression is performed to evaluate the posterior mean function $\overline{g}_j(\tau|g_j^1,g_j^2,\dots,g_j^K)$
\begin{equation}
\label{11}
\overline{g}_j(\tau)=\mathbf{d}^T\mathbf{C}_K^{-1}\mathbf{g}_j^T,
\end{equation}
where $\mathbf{C}_K$ is the matrix of $C(\tau_k,\tau_{k'})=ker(\tau_k,\tau_{k'})+\sigma^2\delta_{kk'}$, $\{\tau_{k(k')}\}_{k(k')=1}^K$ are x-values of $K$ training points and $\sigma^2$ is the noise variance of training points, $\mathbf{d}=[ker(\tau_1,\tau),\dots,ker(\tau_K,\tau)]^T$, $\mathbf{g}_j$ are y-values of $K$ training points. Here the covariance kernel is squared exponential kernel $ker(x,x')=\theta_0\exp\left(-\frac{\theta_1}{2}\|x-x'\|^2\right)$ where $\theta_0$ and $\theta_1$ are hyperparameters of GP. We use $\overline{g}_j(\tau)$ to replace the directly estimated $\mathbf{g}_j$ in NMSE \eqref{7}. By using GP-MRS, the NMSE induced by overfitting of $g_j(\tau)$ can be effectively eliminated when $K$ is large (comparison between Fig.~\ref{fig7} and Fig.~\ref{fig9}). 

It is worth noting that GP-MRS cannot be applied to address the problem of $M$ because a too large $M$ will lead to a sparse sector where the GP regression cannot provide a true posterior mean function. To obtain the optimal hyperparameter $M$, an empirical formula is provided: $M\approx\mathcal{N}/250L$ where $L$ is the number of independent observations. 





\subsection{Synthetic Data Experiment of GP-MRS}

\begin{figure}[htbp]
\centerline{\includegraphics[width=0.45\textwidth]{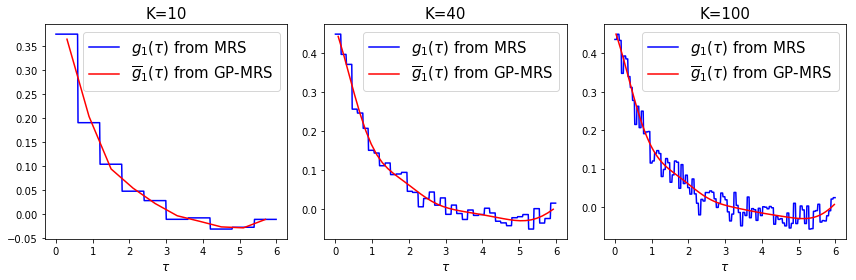}}
\caption{Given $M=10$, the estimated $g_1(\tau)$ from MRS and GP-MRS when $K=10,40$ and $100$.}
\label{fig8}
\end{figure}

\begin{figure}[htbp]
\centerline{\includegraphics[width=0.5\textwidth]{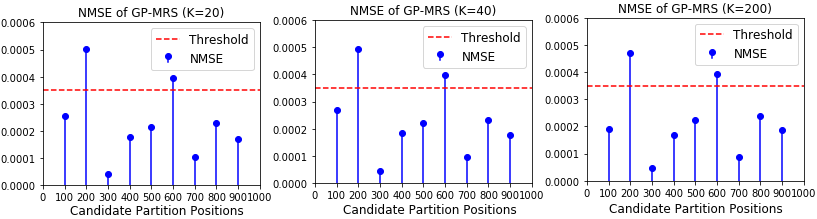}}
\caption{Given $M=10$, the NMSE of GP-MRS w.r.t. $K$. The threshold corresponds to $R=3$ (Only $K=20,40,200$ are shown).The NMSE is nearly unchanged whatever $K$ is.}
\label{fig9}
\end{figure}

We apply the GP-MRS algorithm to the same experiment as in the section \ref{db}. The GP hyperparameters are set to $\theta_0=1, \theta_1=1, \sigma^2=0.01$. It is out of the scope of this paper to discuss how to choose the GP hyperparameters. The estimated $\overline{g}_1(\tau)$ when $K=10,40$ and $100$ is shown in Fig.~\ref{fig8}. It is clear that the $\overline{g}_j(\tau)$ from GP-MRS is stable whatever $K$ is. We also analyze the segmentation and NMSE results with $R=3$ which are shown in Tab.~\ref{tab5} and Fig.~\ref{fig9}; we can see the segmentation and NMSE are both stable whatever $K$ is. \textbf{Conclusively, the GP-MRS can provide the correct segmentation in cases where the MRS does not work}. 

\begin{table}
\centering
\scalebox{0.8}{
\begin{tabular}{|c|c|c|c|c|}
\hline
\textbf{$K$} & 20 & 30  & 40 & 200   \\ \hline
\textbf{Partition Positions} & 200,600 & 200,600 & 200,600  & 200,600 \\ \hline
\end{tabular}}
\caption{Segmentation Results of GP-MRS w.r.t. $K$}
\label{tab5}
\end{table}

\subsection{Computation Complexity of GP-MRS}

For a standard GP, it costs $\mathcal{O}(K^3)$ for complexity when calculating $K$ training points ($\mathbf{g}_j=[g_j^k]_{k=1}^K$). The final complexity of GP-MRS is $\mathcal{O}(\mathcal{N}K+MK^3)$. Unavoidably, the introduce of GP regression into MRS will make it slower. Given $\mathcal{N}=120,000$ and $M=10$, the consuming time of GP-MRS when $K=40$ is 48.64 seconds on a normal desktop. We can see it is still acceptable when $K$ is not too large. 

\section{Real Data Experiment}
\label{sec7}

The GP-MRS is applied to a real vehicle collision dataset to discover the hierarchical time-varying characteristics. 

\subsection{Vehicle Collisions in New York City}

The vehicle collision dataset is provided by the New York City Police Department. It contains about 1.05 million vehicle collision records in New York City from July, 2012 to September, 2017. The dataset includes the collision date, time, borough, location, contributing factor and so on. 

In daily transportation, the vehicle collision occurring in the past will increase the intensity of vehicle collision occurring in the future because of the traffic jam caused by the initial collision, so there exists a triggering effect from the past collision to the future one. There are already some works trying to model the triggering effect using classic Hawkes process (parametric or nonparametric), but they all assume the stationarity is satisfied. However, this is not the case in real life. As shown later, we reveal the hierarchical time-varying characteristics of triggering kernel and baseline intensity of vehicle collision over 24 hours by using GP-MRS algorithm. 

\subsubsection{Weekdays}

We filter out the collision records on all weekdays from May 1st 2017 to June 30th 2017. There are some collisions occurring at the same time as the data resolution is at minute level, which violates the definition of the temporal point process. To avoid this, we add a small time interval to all the simultaneous records to separate them. 

The observation every day is assumed to be independent, so there are 45 sets of independent observations. Totally, 137,578 points are observed. We use the GP-MRS for segmentation which is still fast enough in this case. The whole observation period $T$ is set to 1440 minutes (24 hours a day). The support of $\phi(\tau)$ is set to 8 minutes. The hyperparameters of GP-MRS $\theta_0$, $\theta_1$, $\sigma^2$ are set to 1, 1, 0.01; $K$ is arbitrarily set to 20 and $M$ is set to 12 by using the empirical formula which means the sector size is 120 minutes (2 hours). 

\begin{figure}[htbp]
\centerline{\includegraphics[width=0.38\textwidth]{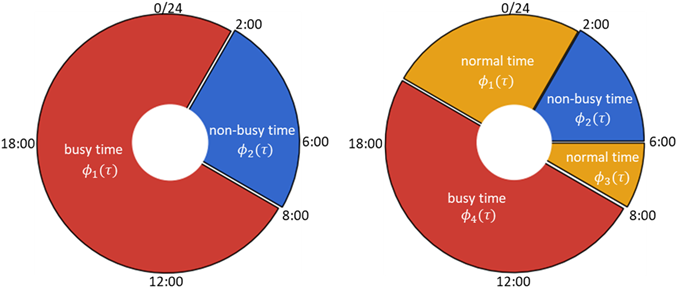}}
\caption{Weekdays: The 24-hour segmentation result of vehicle collisions, 2 segments (left) and 4 segments (right).}
\label{fig10}
\end{figure}

\begin{figure}[htbp]
\centerline{\includegraphics[width=0.4\textwidth]{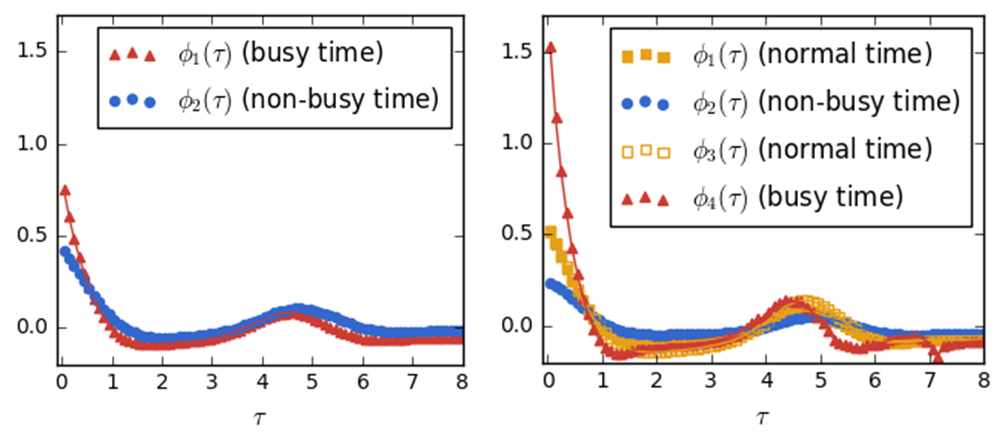}}
\caption{Weekdays: The estimated $\phi(\tau)$ of vehicle collisions, 2 segments (left) and 4 segments (right).}
\label{fig11}
\end{figure}

When the desired output resolution $R=2$, the consuming time of GP-MRS is about 10 seconds and the cutting positions are 2:00 and 8:00. The segmentation is shown in Fig.~\ref{fig10} left and can be understood as the busy time and non-busy time. 
After segmentation, we estimate $\mu$ and $\phi(\tau)$ on each segment. The estimated $\mu$'s are $\mu_1=0.317$ and $\mu_2=0.127$, the estimated $\phi(\tau)$'s are shown in Fig.~\ref{fig11} left. We can see both $\mu_1$ and $\phi_1(\tau)$ are larger than $\mu_2$ and $\phi_2(\tau)$ which is consistent with our common sense because the traffic is more crowd in busy time. Additionally, the nonparametricly estimated triggering kernel is not strictly monotonic decreasing: there is a small bump around 5 minutes after the initial collision, which
proves the superior flexibility of nonparametric estimation. 

To show the hierarchical multi-resolution property of GP-MRS, the desired output resolution $R$ is increased to 4 and we obtain a finer segmentation. The consuming time in this case is also about 10 seconds and the cutting positions are 2:00, 6:00, 8:00 and 20:00. The segmentation is shown in Fig.~\ref{fig10} right. The segmentation can be understood as the normal time, busy time and non-busy time. The late night is between 2:00 and 6:00 which are non-busy hours; the after-work entertainment hours (from 20:00 to 2:00) together with morning commute hours (from 6:00 to 8:00) are the normal time; the daytime (from 8:00 to 20:00) is the busy time. The estimated $\mu$'s are $\mu_1=0.32$, $\mu_2=0.12$, $\mu_3=0.29$ and $\mu_4=0.59$. The estimated $\phi(\tau)$'s are shown in Fig.~\ref{fig11} right. Two normal-time $\phi(\tau)$'s are almost overlapping; both the baseline intensity and triggering kernel of busy time are larger than normal time, larger than non-busy time at the initial stage. 


\subsubsection{Weekends}

\begin{figure}[htbp]
\centerline{\includegraphics[width=0.38\textwidth]{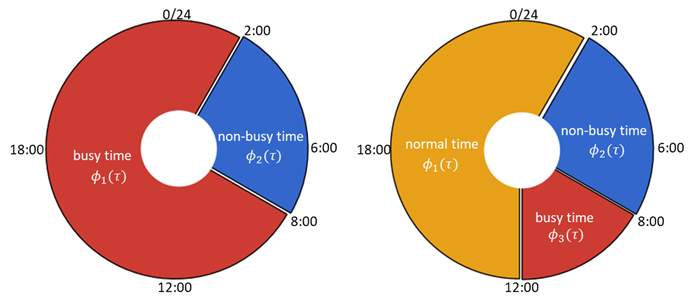}}
\caption{Weekends: The 24-hour segmentation result of vehicle collisions, 2 segments (left) and 3 segments (right).}
\label{fig13}
\end{figure}

\begin{figure}[htbp]
\centerline{\includegraphics[width=0.4\textwidth]{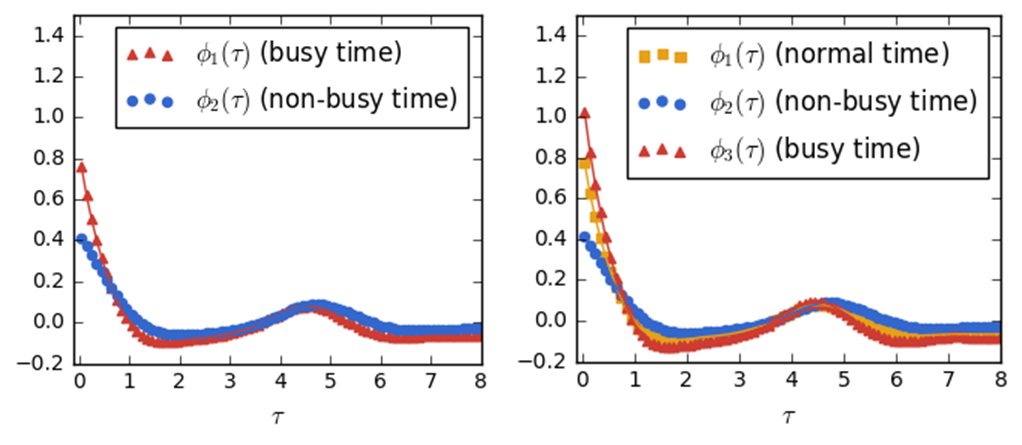}}
\caption{Weekends: The estimated $\phi(\tau)$ of vehicle collisions, 2 segments (left) and 3 segments (right).}
\label{fig14}
\end{figure}

We also filter out collision records on all weekends from February 1st 2017 to August 31th 2017. With $R=2$, the cutting positions are 2:00 and 8:00 which are same as weekdays. With $R=3$, we can get a finer segmentation: 2:00, 8:00 and 12:00. The segmentation is shown in Fig.~\ref{fig13}. The estimated $\phi(\tau)$'s are shown in Fig.~\ref{fig14}. 

An interesting phenomenon is that \textbf{the low-resolution time-varying characteristics of weekdays are similar with that of weekends, but the high-resolution characteristics are very different}, e.g. 6:00-8:00 becomes the non-busy time on weekends maybe because of late waking up; 12:00-20:00 becomes the normal time maybe because of less heavy traffic. \textbf{The multi-resolution segmentation provides a hierarchical insight into the dynamic evolution of vehicle collision}. 

\subsection{Comparison with Other Models}

To show the superiority of our model, we compare the learned results of stationary parametric (vanilla version), stationary nonparametric (nonparametric version) and nonstationary nonparametric Hawkes process (our proposed version). 

\begin{figure}[htbp]
\centerline{\includegraphics[width=0.38\textwidth]{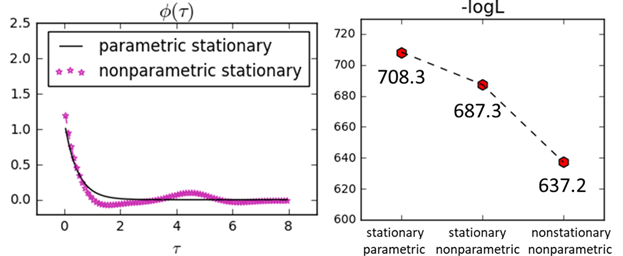}}
\caption{The estimated $\phi(\tau)$ of vehicle collisions of stationary parametric and stationary nonparametric Hawkes process (left). The test error ($-logL$) of three models (right).}
\label{fig15}
\end{figure}

For stationary parametric version, we assume the triggering kernel is an exponential decay function: $\phi(\tau)=\alpha\cdot\exp(-\beta\tau)$. The inference is performed by maximum likelihood estimation (MLE). The learned $\mu=0.22$ and $\phi(\tau)$ is shown in Fig.~\ref{fig15}. For stationary nonparametric version, there is only one triggering kernel which is nonparametric. The inference is performed by Wiener-Hopf equation method. The learned $\mu=0.26$ and $\phi(\tau)$ is shown in Fig.\ref{fig15}. 

We compare the test error (negative log-likelihood $-logL$) of three models on the vehicle collision test data (weekdays). The result is shown in Fig.\ref{fig15} and our proposed model (the nonstationary nonparametric version) fits the data best. 

\section{Conclusions}
\label{sec8}

In this paper, we propose the first MRS algorithm to partition the nonstationary Hawkes process, which provides a hierarchical view of the nonstationary structure. By this way, the hierarchical dynamic time-varying characteristics of nonstaionary Hawkes process can be discovered. Besides, the algorithm is fast because of the use of cumulants. After segmentation, the baseline intensity and triggering kernel are estimated in a nonparametric way. Overall, this is a nonstationary and nonparametric Hawkes process. To ease the choice of hyperparameter $K$, the GP-MRS algorithm is also proposed at the cost of lower efficiency but still acceptable. Both synthetic and real data experiments show the superiority of our proposed model.

\bibliographystyle{named}
\bibliography{ijcai19}

\begin{thebibliography}{}

\bibitem[\protect\citeauthoryear{Atkinson}{1976}]{atkinson1976survey}
Kendall Atkinson.
\newblock A survey of numerical methods for the solution of {F}redholm integral
  equations of the second kind.
\newblock 1976.

\bibitem[\protect\citeauthoryear{Bacry and Muzy}{2016}]{bacry2016first}
Emmanuel Bacry and Jean-Fran{\c{c}}ois Muzy.
\newblock First-and second-order statistics characterization of {H}awkes
  processes and non-parametric estimation.
\newblock {\em IEEE Transactions on Information Theory}, 62(4):2184--2202,
  2016.

\bibitem[\protect\citeauthoryear{Bernaola-Galv{\'a}n \bgroup \em et al.\egroup
  }{2001}]{bernaola2001scale}
Pedro Bernaola-Galv{\'a}n, Plamen~Ch Ivanov, Lu{\'\i}s A~Nunes Amaral, and
  H~Eugene Stanley.
\newblock Scale invariance in the nonstationarity of human heart rate.
\newblock {\em Physical review letters}, 87(16):168105, 2001.

\bibitem[\protect\citeauthoryear{Bernaola-Galv{\'a}n \bgroup \em et al.\egroup
  }{2012}]{bernaola2012segmentation}
P~Bernaola-Galv{\'a}n, JL~Oliver, M~Hackenberg, AV~Coronado, P~Ch Ivanov, and
  P~Carpena.
\newblock Segmentation of time series with long-range fractal correlations.
\newblock {\em The European Physical Journal B}, 85(6):211, 2012.

\bibitem[\protect\citeauthoryear{Carlstein \bgroup \em et al.\egroup
  }{1994}]{carlstein1994change}
Edward~G Carlstein, Hans-Georg M{\"u}ller, and David Siegmund.
\newblock Change-point problems.
\newblock IMS, 1994.

\bibitem[\protect\citeauthoryear{Du \bgroup \em et al.\egroup
  }{2016}]{du2016recurrent}
Nan Du, Hanjun Dai, Rakshit Trivedi, Utkarsh Upadhyay, Manuel Gomez-Rodriguez,
  and Le~Song.
\newblock Recurrent marked temporal point processes: {E}mbedding event history
  to vector.
\newblock In {\em Proceedings of the 22nd ACM SIGKDD International Conference
  on Knowledge Discovery and Data Mining}, pages 1555--1564. ACM, 2016.

\bibitem[\protect\citeauthoryear{Feng \bgroup \em et al.\egroup
  }{2005}]{feng2005abrupt}
Guo-Ling Feng, Zhi-Qiang Gong, Wen-Jie Dong, and Jian-Ping Li.
\newblock Abrupt climate change detection based on heuristic segmentation
  algorithm.
\newblock 2005.

\bibitem[\protect\citeauthoryear{Gupta \bgroup \em et al.\egroup
  }{2018}]{gupta2018discrete}
Amrita Gupta, Mehrdad Farajtabar, Bistra Dilkina, and Hongyuan Zha.
\newblock Discrete interventions in {H}awkes processes with applications in
  invasive species management.
\newblock In {\em IJCAI}, pages 3385--3392, 2018.

\bibitem[\protect\citeauthoryear{Hawkes}{1971}]{hawkes1971spectra}
Alan~G Hawkes.
\newblock Spectra of some self-exciting and mutually exciting point processes.
\newblock {\em Biometrika}, 58(1):83--90, 1971.

\bibitem[\protect\citeauthoryear{Jovanovi{\'c} \bgroup \em et al.\egroup
  }{2015}]{jovanovic2015cumulants}
Stojan Jovanovi{\'c}, John Hertz, and Stefan Rotter.
\newblock Cumulants of {H}awkes point processes.
\newblock {\em Physical Review E}, 91(4):042802, 2015.

\bibitem[\protect\citeauthoryear{Lemonnier and
  Vayatis}{2014}]{lemonnier2014nonparametric}
Remi Lemonnier and Nicolas Vayatis.
\newblock Nonparametric markovian learning of triggering kernels for mutually
  exciting and mutually inhibiting multivariate {H}awkes processes.
\newblock In {\em Joint European Conference on Machine Learning and Knowledge
  Discovery in Databases}, pages 161--176. Springer, 2014.

\bibitem[\protect\citeauthoryear{Lewis and
  Mohler}{2011}]{lewis2011nonparametric}
Erik Lewis and George Mohler.
\newblock A nonparametric {E}{M} algorithm for multiscale {H}awkes processes.
\newblock {\em Journal of Nonparametric Statistics}, 1(1):1--20, 2011.

\bibitem[\protect\citeauthoryear{Liu \bgroup \em et al.\egroup
  }{2018}]{liu2018exploiting}
Yanchi Liu, Tan Yan, and Haifeng Chen.
\newblock Exploiting graph regularized multi-dimensional {H}awkes processes for
  modeling events with spatio-temporal characteristics.
\newblock In {\em IJCAI}, pages 2475--2482, 2018.

\bibitem[\protect\citeauthoryear{Luo \bgroup \em et al.\egroup
  }{2015}]{luo2015multi}
Dixin Luo, Hongteng Xu, Yi~Zhen, Xia Ning, Hongyuan Zha, Xiaokang Yang, and
  Wenjun Zhang.
\newblock Multi-task multi-dimensional {H}awkes processes for modeling event
  sequences.
\newblock In {\em Twenty-Fourth International Joint Conference on Artificial
  Intelligence}, 2015.

\bibitem[\protect\citeauthoryear{Noble and Weiss}{1959}]{noble1959methods}
Benjamin Noble and George Weiss.
\newblock Methods based on the {W}iener-{H}opf technique for the solution of
  partial differential equations.
\newblock {\em Physics Today}, 12:50, 1959.

\bibitem[\protect\citeauthoryear{Nystr{\"o}m}{1930}]{nystrom1930praktische}
Evert~J Nystr{\"o}m.
\newblock {\"U}ber die praktische {A}ufl{\"o}sung von {I}ntegralgleichungen mit
  {A}nwendungen auf {R}andwertaufgaben.
\newblock {\em Acta Mathematica}, 54(1):185--204, 1930.

\bibitem[\protect\citeauthoryear{Ogata}{1998}]{ogata1998space}
Yosihiko Ogata.
\newblock Space-time point-process models for earthquake occurrences.
\newblock {\em Annals of the Institute of Statistical Mathematics},
  50(2):379--402, 1998.

\bibitem[\protect\citeauthoryear{Roueff and Von~Sachs}{2017}]{roueff2017time}
Fran{\c{c}}ois Roueff and Rainer Von~Sachs.
\newblock Time-frequency analysis of locally stationary {H}awkes processes.
\newblock {\em arXiv preprint arXiv:1704.01437}, 2017.

\bibitem[\protect\citeauthoryear{Roueff \bgroup \em et al.\egroup
  }{2016}]{roueff2016locally}
Fran{\c{c}}ois Roueff, Rainer Von~Sachs, and Laure Sansonnet.
\newblock Locally stationary {H}awkes processes.
\newblock {\em Stochastic Processes and their Applications}, 126(6):1710--1743,
  2016.

\bibitem[\protect\citeauthoryear{Tannenbaum and
  Burak}{2017}]{tannenbaum2017theory}
Neta~Ravid Tannenbaum and Yoram Burak.
\newblock Theory of nonstationary {H}awkes processes.
\newblock {\em Physical Review E}, 96(6):062314, 2017.

\bibitem[\protect\citeauthoryear{Thompson}{2012}]{thompson2012point}
W~Thompson.
\newblock {\em Point process models with applications to safety and
  reliability}.
\newblock Springer Science \& Business Media, 2012.

\bibitem[\protect\citeauthoryear{Toth \bgroup \em et al.\egroup
  }{2010}]{toth2010segmentation}
Bence Toth, Fabrizio Lillo, and J~Doyne Farmer.
\newblock Segmentation algorithm for non-stationary compound {P}oisson
  processes.
\newblock {\em The European Physical Journal B}, 78(2):235--243, 2010.

\bibitem[\protect\citeauthoryear{Weinberg \bgroup \em et al.\egroup
  }{2007}]{weinberg2007bayesian}
Jonathan Weinberg, Lawrence~D Brown, and Jonathan~R Stroud.
\newblock Bayesian forecasting of an inhomogeneous {P}oisson process with
  applications to call center data.
\newblock {\em Journal of the American Statistical Association},
  102(480):1185--1198, 2007.

\end{thebibliography}

\end{document}